\pdfoutput=1
\documentclass[prd,aps,twocolumn,preprintnumbers,amsmath,amssymb,superscriptaddress]{revtex4}

\usepackage[dvips]{graphicx}

\usepackage{dcolumn}
\usepackage{bm}
\usepackage{color}

\begin{document}


\title{Complementarity of Dark Matter Direct Detection Targets}

\author{Miguel Pato}
\email{pato@iap.fr}
\affiliation{Institute for Theoretical Physics, Univ.~of Z\"urich, Winterthurerst.~190, 8057 Z\"urich CH}
\affiliation{Institut d'Astrophysique de Paris, UMR 7095-CNRS, Univ.~Pierre \& Marie Curie, 98bis Bd Arago 75014 Paris, France}
\affiliation{Dipartimento di Fisica, Universit\`a degli Studi di Padova, via Marzolo 8, I-35131, Padova, Italy}
\author{Laura Baudis}
\affiliation{Physics Institute, Univ.~of Z\"urich, Winterthurerst.~190, 8057 Z\"urich CH}
\author{Gianfranco Bertone}
\affiliation{Institute for Theoretical Physics, Univ.~of Z\"urich, Winterthurerst.~190, 8057 Z\"urich CH}
\affiliation{Institut d'Astrophysique de Paris, UMR 7095-CNRS, Univ.~Pierre \& Marie Curie, 98bis Bd Arago 75014 Paris, France}
\author{Roberto Ruiz de Austri}
\affiliation{Instituto de F\'isica Corpuscular, IFIC-UV/CSIC, Valencia, Spain} 
\author{Louis E. Strigari}
\affiliation{Kavli Institue for Particle Astrophysics \& Cosmology, Stanford University, Stanford, CA, 94305}
\author{Roberto Trotta}
\affiliation{Astrophysics Group, Imperial College London \\
	Blackett Laboratory, Prince Consort Road, London SW7 2AZ, UK}

\date{\today}

\begin{abstract}
We investigate the reconstruction capabilities of Dark Matter mass and spin-independent cross-section from future ton-scale direct detection experiments using germanium, xenon or argon as targets. Adopting realistic values for the exposure, energy threshold and resolution of Dark Matter experiments which will come online within 5 to 10 years, the degree of complementarity between different targets is quantified. We investigate how the uncertainty in the astrophysical parameters controlling the local Dark Matter density and velocity distribution affects the reconstruction. For a 50 GeV WIMP, astrophysical uncertainties degrade the accuracy in the mass reconstruction by up to a factor of $\sim 4$ for xenon and germanium, compared to the case when astrophysical quantities are fixed. However, combination of argon, germanium and xenon data increases the constraining power by a factor of $\sim 2$ compared to germanium or xenon alone. We show that future direct detection experiments can achieve self-calibration of some astrophysical parameters, and they will be able to constrain the WIMP mass with only very weak external astrophysical constraints.
\end{abstract}

\maketitle

\section{Introduction}\label{secintro}

\par Many experiments are currently searching for Dark Matter (DM) in the form of Weakly Interacting Massive Particles (WIMPs), by looking for rare scattering events off nuclei in the detectors, and many others are planned for the next decade \cite{book,Lewin,Bergstrom,Munoz,Bertone,CerdenoGreen}. This direct DM detection strategy has brought over the last year several interesting observations and upper limits. The results of the DAMA/LIBRA \cite{dama} and, more recently, the CoGeNT \cite{cogent} collaborations have been tentatively interpreted as due to DM particles. It appears however that these results cannot be fully reconciled with other experimental findings, in particular with the null searches from XENON100 \cite{xe100,gelmini1,Bezrukov} or CDMS \cite{cdms10}, and are also in tension with ZEPLIN-III \cite{zeplin}. In this context, the next generation of low-background, underground detectors is eagerly awaited and will hopefully confirm or rule out a DM interpretation.
 
\par If convincing evidence is obtained for DM particles with direct detection experiments, the obvious next step will be to attempt a reconstruction of the physical parameters of the DM particle, namely its mass and scattering cross-section (see e.g.~Refs.~\cite{Goudelis,Green1,ST}). This is a non-trivial task, hindered by the different uncertainties associated with the computation of WIMP-induced recoil spectra. In particular, Galactic model uncertainties -- i.e.~uncertainties pertaining to the density and velocity distribution of WIMPs in our neighbourhood -- play a crucial role. In attempting reconstruction, the simplest assumption to make is a fixed local DM density $\rho_0=0.3 \textrm{ GeV/cm}^3$ and a ``standard halo model'', i.e.~an isotropic isothermal sphere density profile and a Maxwell-Boltzmann distribution of velocities with a given galactic escape velocity $v_{esc}$ and one-dimensional dispersion $\sigma^2\equiv v_0^2/2 = v_{lsr}^2/2$ ($v_0$ being the most probable velocity and $v_{lsr}$ the local circular velocity, see below). However, the Galactic model parameters are only estimated to varying degrees of accuracy, so that the true local population of DM likely deviates from the highly idealised standard halo model.

\par Several attempts have been made to improve on the standard approach \cite{ST,Ling,McCabe,Green2}. In the case of a detected signal at one experiment, recent analyses have studied how complementary detectors can extract dark matter properties, independent of our knowledge of the Galactic model~\cite{fox1}. Certain properties of dark matter may also be extracted under assumptions about the nature of the nuclear recoil events~\cite{fox2}. Furthermore, eventual multiple signals at different targets have been shown to be useful in constraining both dark matter and astrophysical properties \cite{APeter} and in extracting spin-dependent and spin-independent couplings \cite{Vergados,BertoneCerdeno}. Here, using a Bayesian approach, we study how uncertainties on Galactic model parameters affect the determination of the DM mass $m_{\chi}$ and spin-independent WIMP-proton scattering cross-section $\sigma_{SI}^p$. In particular we focus on realistic experimental capabilities for the future generation of ton-scale detectors -- to be reached within the next 10 years -- with noble liquids (argon, xenon) and cryogenic (germanium) technologies.

\par The main focus of this paper is the complementarity between different detection targets. It is well-known (see e.g.~\cite{Lewin}) that different targets are sensitive to different directions in the $m_{\chi}-\sigma_{SI}^p$ plane, which is very useful to achieve improved reconstruction capabilities -- or more stringent bounds in the case of null results. This problem has often been addressed without taking proper account of Galactic model uncertainties. Using xenon (Xe), argon (Ar) and germanium (Ge) as case-studies, we ascertain to what extent unknowns in Galactic model parameters limit target complementarity. A thorough understanding of complementarity will be crucial in the near future since it provides us with a sound handle to compare experiments and, if needed, decide upon the best target to bet on future detectors. Our results also have important consequences for the combination of collider observables and direct detection results (for a recent work see \cite{Fornasa}).

\par Besides degrading the extraction of physical properties like $m_{\chi}$ and $\sigma_{SI}^p$, uncertainties in the Galactic model will challenge our ability to distinguish between different particle physics frameworks in case of a positive signal. Other relevant unknowns are hadronic uncertainties, related essentially to the content of nucleons \cite{Ellis}. Here, we undertake a model-independent approach without specifying an underlying WIMP theory and using $m_{\chi}$ and $\sigma_{SI}^p$ as our phenomenological parameters -- for this reason we shall not address hadronic uncertainties (hidden in $\sigma_{SI}^p$). A comprehensive work complementary to ours and done in the supersymmetric framework has been presented recently \cite{Akrami1,Akrami2}.

\par The paper is organised as follows. In the next section, we give some basic formulae for WIMP-nucleus recoil rates in direct detection experiments. In Section \ref{exp} the upcoming experimental capabilities are detailed, while Section \ref{statsection} describes our Bayesian approach. We outline the relevant Galactic model uncertainties and our modelling of the velocity distribution function in Section \ref{astrounc} and present our results in Section \ref{results} before concluding in Section \ref{secconc}.

\section{Basics of direct dark matter detection}\label{DD}

\par Several thorough reviews on direct dark matter searches exist in the literature \cite{book,Lewin,Bergstrom,Munoz,Bertone,CerdenoGreen}. In this section, we simply recall the relevant formulae,  emphasizing the impact of target properties and unknown quantities.

\par The elastic recoil spectrum produced by WIMPs of mass $m_{\chi}$ and local density $\rho_0$ on target nuclei $N(A,Z)$ of mass $m_N$ is 
\begin{equation}\label{recoil1}
 \frac{dR}{dE_R}(E_R)=\frac{\rho_0}{m_{\chi}m_{N}} \, \int_{\cal V}
 { d^3\vec{v} \, \, v f\left(\vec{v}+\vec{v_e}\right)  \frac{d\sigma_{\chi-N}}{dE_R}(v,E_R) } ,
\end{equation}
where $\vec{v}$ is the WIMP velocity in the detector rest frame, $\vec{v}_e$ is the Earth velocity in the Galactic rest frame, $f(\vec{w})$ is the WIMP velocity distribution in the galactic rest frame and $\sigma_{\chi-N}$ is the WIMP-nucleus cross-section. The integral is performed over ${\cal V}: v>v_{min}(E_R)$, where $v_{min}$ is the minimum WIMP velocity that produces a nuclear recoil of energy $E_R$. Eq.~\eqref{recoil1} simply states that the recoil rate is the flux of WIMPs $\rho_0 v / m_{\chi}$, averaged over the velocity distribution $f(\vec{w})$, times the probability of interaction with one target nucleus $\sigma_{\chi-N}$. Anticipating the scale of future detectors, we will think of measuring $dR/dE_R$ in units of counts/ton/yr/keV. For non-relativistic (elastic) collisions -- as appropriate for halo WIMPs, presenting $v/c\sim 10^{-3}$ -- the kinematics fixes the recoil energy
\begin{equation*}
E_R(m_{\chi},v,A,\theta')=\frac{\mu_N^2 v^2 (1-\cos\theta')}{m_N} \quad ,
\end{equation*}  
and the minimum velocity
\begin{equation*}
v_{min}(m_{\chi},E_R,A)=\sqrt{\frac{m_N E_R}{2 \mu_N^2}}
\end{equation*}
in which $\theta'$ is the scattering angle in the centre of mass and $\mu_N=\frac{m_{\chi}m_N}{m_{\chi}+m_N}$ is the WIMP-nucleus reduced mass. 

\par In principle, all WIMP-nucleus couplings enter in the cross-section $\sigma_{\chi-N}$. However, we shall focus solely on spin-independent (SI) scalar interactions so that 
\begin{equation*}
\frac{d\sigma_{\chi-N}}{dE_R}=\frac{m_N}{2 \mu_N^2 v^2} \sigma_{SI}^N F^2(A,E_R) \quad ,
\end{equation*}
where $\sigma_{SI}^N = \frac{4 \mu_N^2}{\pi} \left[ Z f_p + (A-Z) f_n \right]^2$ is the WIMP-nucleus spin-independent cross-section at zero momentum transfer and $F(A,E_R)$ is the so-called form factor that accounts for the exchange of momentum. Assuming that the WIMP couplings to protons and neutrons are similar, $f_p\sim f_n$, and defining $\sigma_{SI}^p\equiv \frac{4\mu_p^2}{\pi}f_p^2$ ($\mu_p$ being the WIMP-proton reduced mass), one gets
\begin{equation}\label{sigma}
\frac{d\sigma_{\chi-N}}{dE_R} = \frac{m_N}{2\mu_p^2 v^2} \sigma_{SI}^p A^2 F^2(A,E_R) \quad . 
\end{equation}
For the form factor, we use the parameterisation in \cite{Lewin} appropriate for spin-independent couplings, namely 
\begin{equation*}
F(A,E_R)=3\frac{\textrm{sin}(q r_n)-(q r_n)\textrm{cos}(q r_n)}{(q r_n)^3} \textrm{exp}(-(q s)^2/2) \quad ,
\end{equation*}
with $qr=6.92\times 10^{-3} A^{1/2} (E/\textrm{keV})^{1/2} r/\textrm{fm}$, $s\simeq 0.9$ fm, $r_n^2=c^2+\frac{7}{3}\pi^2 a^2-5 s^2$, $c/\textrm{fm}=1.23A^{1/3}-0.6$ and $a\simeq0.52$ fm.

\par As noticed above, in Eq.~\eqref{recoil1} $\vec{v}_e$ is the Earth velocity with respect to the galactic rest frame and amounts to $\vec{v}_e=\vec{v}_{lsr}+\vec{v}_{pec}+\vec{v}_{orb}$, where $v_{lsr}\sim\mathcal{O}(250)$ km/s is the local circular velocity, $v_{pec}\sim\mathcal{O}(10)$ km/s is the peculiar velocity of the Sun (with respect to $\vec{v}_{lsr}$) and $v_{orb}\sim\mathcal{O}(30)$ km/s is the Earth velocity with respect to the Sun (i.e.~the Earth orbit). Here, we are not interested in the annual modulation signal nor directional signatures but rather in the average recoil rate -- therefore we shall neglect $\vec{v}_{pec}$ and $\vec{v}_{orb}$ and take $\vec{v}_e\simeq \vec{v}_{lsr} = \text{const}$. 

\par Under these assumptions, Eq.~\eqref{recoil1} may be recast in a very convenient way:
\begin{eqnarray}\nonumber
\frac{dR}{dE_R}(E_R)&=&\frac{\rho_0 \sigma_{SI}^p}{2 \mu_p^2 m_{\chi}} \times A^2 F^2(A,E_R) \times \\ \label{recoil2}
 & & \mathcal{F}\left(v_{min}(m_{\chi},E_R,A),\vec{v}_e;v_0,v_{esc}\right) \, ,
\end{eqnarray}
where we have used Eq.~\eqref{sigma}, defined 
\begin{equation}
\label{eq:F}
\mathcal{F}\equiv\int_{\cal V}{ d^3\vec{v} \frac{f\left(\vec{v}+\vec{v_e}\right)}{v}}
\end{equation} 
and made explicit the dependence of $\mathcal{F}$ on the velocity distribution parameters $v_0$ and $v_{esc}$. Below we discuss in more detail the connection between the parameters $v_0$ and $v_{lsr}$. The distribution of DM is encoded in the factor $\mathcal{F}$ (and $\rho_0$), whereas the detector-related quantities appear in $A^2 F^2(A,E_R)$ (and $v_{min}$). The apparent degeneracy along the direction $\rho_0 \sigma_{SI}^p/m_{\chi}=\text{const}$ may be broken by using different recoil energies and/or different targets since $\mathcal{F}$ is sensitive to a non-trivial combination of $m_{\chi}$, $E_{R}$ and $A$. Nevertheless, for very massive WIMPs $m_{\chi}\gg m_{N}\sim \mathcal{O}(100) \textrm{ GeV} \gg m_p$, the minimum velocity becomes independent of $m_{\chi}$, $v_{min}\simeq \sqrt{E_R/(2m_N)}$, and the degeneracy $\rho_0 \sigma_{SI}^p/m_{\chi}$ cannot be broken. Depending on the target being used, this usually happens for WIMP masses above a few hundred GeV.

\par Ultimately, the observable we will be interested in is the number of recoil events in a given energy bin $E_1 < E_R < E_2$:
\begin{equation} \label{eq:NR}
N_R(E_1,E_2) = \int_{E_1}^{E_2}{ dE_R \, \, \epsilon_{eff} \, \frac{d\tilde{R}}{dE_R} } \quad ,
\end{equation}
$\epsilon_{eff}$ being the effective exposure (usually expressed in ton$\times$yr) and $d\tilde{R}/dE_R$ the recoil rate smeared according to the energy resolution of the detector $\sigma(E)$,
\begin{equation*}
\frac{d\tilde{R}}{dE_R} = \int{ dE' \, \frac{dR}{dE_R}(E') \, \frac{1}{\sqrt{2\pi}\sigma(E')} \textrm{exp}\left(-\frac{(E-E')^2}{2\sigma^2(E')}\right)  } \quad .
\end{equation*}

\par Three fiducial WIMP models will be used to assess the capabilities of future direct detection experiments: $m_{\chi}=$25, 50 and 250 GeV, all with $\sigma_{SI}^p=10^{-9}$ pb. These models are representative of well-motivated candidates such as neutralinos in supersymmetric theories \cite{ruiz}.

\section{Upcoming experimental capabilities}\label{exp}

\begin{table*}[htp]
\centering
\fontsize{9}{9}\selectfont
\begin{tabular}{c|ccccccc}
\hline
\hline
 target & $\epsilon$ [ton$\times$yr] & $\eta_{cut}$ & $A_{NR}$ & $\epsilon_{eff}$ [ton$\times$yr] & $E_{thr}$ [keV] & $\sigma(E)$ [keV]  &  background events/$\epsilon_{eff}$ \\
\hline
Xe & 5.0 & 0.8 & 0.5 & 2.00 & 10 & Eq.~\eqref{resXe} &  $<1$ \\
Ge & 3.0 & 0.8 & 0.9 & 2.16 & 10 & Eq.~\eqref{resGe} & $<1$ \\
Ar & 10.0 &0.8 & 0.8 & 6.40  & 30 & Eq.~\eqref{resAr} &  $<1$ \\
\hline
\end{tabular}
\caption{\fontsize{9}{9}\selectfont Characteristics of future direct dark matter experiments using xenon, germanium and argon as target nuclei. In all cases the level of background in the fiducial mass region is negligible for the corresponding effective exposure. See Section \ref{exp} for further details.}\label{tabExp}
\end{table*}

\par Currently, the most stringent constraints on the SI WIMP-nucleon coupling are those obtained by the CDMS \cite{cdms} and XENON \cite{xe100} collaborations. While XENON100 should probe the cross-section region down to  $5\times10^{-45}$ cm$^2$ with data already in hand, the XENON1T \cite{xenon1T} detector, whose construction is scheduled to start by mid 2011, is expected to reach another order of magnitude in sensitivity improvement. 
To test the  $\sigma_{SI}^p$  region down to $10^{-47}$ cm$^2\equiv 10^{-11}$ pb and below, a new generation of detectors with larger WIMP target masses and ultra-low backgrounds is needed.   Since we are interested in the prospects for detection in the next 5 to 10 years, we discuss new projects that can realistically be built on this time scale, adopting the most promising detection techniques, namely noble liquid time projection chambers (TPCs) and cryogenic detectors operated at mK temperatures. 

In Europe, two large consortia, DARWIN \cite{darwin} and EURECA \cite{eureca}, gathering the expertise of several groups working on existing DM experiments  are funded for R\&D and design studies to push noble liquid and cryogenic experiments to the multi-ton and ton scale, respectively. DARWIN is devoted to noble liquids, having as main goal the construction of a multi-ton liquid Xe (LXe) and/or liquid Ar (LAr) instrument \cite{darwin_IDM2010}, with data taking to start around 2016. The XENON, ArDM and WARP collaborations participate actively in the DARWIN project. EURECA is a design study dedicated to cryogenic dark matter detectors operated at mK temperatures. The proposed roadmap is to improve upon CRESST \cite{cresst} and EDELWEISS \cite{edel} technologies and build a ton-scale detector by 2018, with a SI sensitivity of about $10^{-46}$ cm$^2\equiv 10^{-10}$ pb. The complementarity between DARWIN and EURECA is of utmost importance for dark matter direct searches since a solid, uncontroversial discovery requires signals in distinct targets and preferentially distinct technologies. In an international context, two engineering studies (MAX \cite{max} and LZS \cite{lzs}) are funded in the US for ton to multi-ton scale LXe and LAr TPCs and the SuperCDMS/GEODM collaboration  \cite{geodm} plans to operate an 1.5\,ton Ge cryogenic experiment at DUSEL \cite{dusel}. In Japan, the XMASS experiment \cite{xmass}, using a total of 800\,kg of liquid xenon in a single-phase detector, is under commissioning at the Kamioka underground laboratory \cite{kamioka}, while a large single-phase liquid argon detector, DEAP-3600 \cite{deap3600}, using 3.6\,tons of LAr is under construction at SNOLab \cite{snolab}.

\par Given these developments, we will focus on the three most promising targets: Xe and Ar as examples of noble liquid detectors, and Ge as a case-study for the cryogenic technique. In the case of a Ge target, we assume an 1.5 ton detector (1\,ton as fiducial target mass), 3 years of operation, an energy threshold for nuclear recoils of  $E_{thr,Ge}=10$ keV  and an energy resolution given by
\begin{equation}\label{resGe}
\sigma_{Ge}(E)=\sqrt{(0.3)^2+(0.06)^2 E/\textrm{keV}} \textrm{ keV} \quad .
\end{equation}

\par For a liquid Xe detector, we assume a total mass of 8\,tons (5\,tons in the fiducial region), 1 year of operation, an energy threshold for  nuclear recoils of  $E_{thr,Xe}=10$ keV  and an energy resolution of 
\begin{equation}\label{resXe}
\sigma_{Xe}(E) = 0.6 \textrm{ keV} \sqrt{E/\textrm{keV}} \quad .
\end{equation}

\par Finally, for a liquid Ar detector, we assume a total mass of 20\,tons (10\,tons in the fiducial region), 1 year of operation, an energy threshold for  nuclear recoils of  $E_{thr,Ar}=30$ keV  and an energy resolution of \cite{privRegenfuss}
\begin{equation}\label{resAr}
\sigma_{Ar}(E) = 0.7 \textrm{ keV} \sqrt{E/\textrm{keV}} \quad .
\end{equation}

To calculate realistic exposures, we make the following assumptions: nuclear recoils acceptances $A_{NR}$ of 90\%, 80\% and 50\% for Ge, Ar and Xe, respectively, and an additional, overall cut efficiency $\eta_{cut}$ of 80\% in all cases, which for simplicity we consider to be constant in energy. We hypothesise less than one background event per given effective exposure $\epsilon_{eff}$, which amounts to 2.16  ton$\times$yr in Ge, 6.4 ton$\times$yr in Ar and 2 ton$\times$yr in Xe, after allowing for all  cuts. Such an ultra-low background will be achieved by a combination of background rejection using the ratio of charge-to-light in Ar and Xe, and charge-to-phonon in Ge, the timing characteristics of raw signals, the self-shielding properties and extreme radio-purity of detector materials, as well as minimisation of exposure to cosmic rays above ground.

\par The described characteristics are summarised in Table \ref{tabExp}. We note that in the following we shall consider recoil energies below 100 keV only; to increase this maximal value may add some information but the effect is likely small given the exponential nature of WIMP-induced recoiling spectra.

\section{Statistical methodology}\label{statsection}

\newcommand{\params}{\Theta}
\newcommand{\data}{d}
\newcommand{\like}{{\mathcal L}}
\newcommand{\DKL}{D_\text{KL}}

We take a Bayesian approach to parameter inference. We begin by briefly summarizing the basics, and we refer the reader to \cite{Trotta:2008qt} for further details. Bayesian inference rests on Bayes theorem, which reads 
\begin{equation} \label{eq:bayes}
p(\params | \data) = \frac{p(\data | \params) p(\params)}{p(\data)},
\end{equation}
where $p(\params | \data)$ is the posterior probability density function (pdf) for the parameters of interest, $\params$, given data $\data$, $p(\data | \params) = \like(\params)$ is the likelihood function (when viewed as a function of $\params$ for fixed data $\data$) and $p(\params)$ is the prior. Bayes theorem thus updates our prior knowledge about the parameters to the posterior by accounting for the information contained in the likelihood. The normalization constant on the r.h.s.~of Eq.~\eqref{eq:bayes} is the Bayesian evidence and it is given by the average likelihood under the prior:
\begin{equation}
p(\data) = \int d\params p(\data | \params) p(\params).
\end{equation}
The evidence is the central quantity for Bayesian model comparison~\cite{Trotta:2005ar}, but it is just a normalisation constant in the context of the present paper.

\begin{table*}
\centering
\fontsize{9}{9}\selectfont
\begin{tabular}{l | l l}
\hline
\hline
 Parameter & Prior range & Prior constraint \\
\hline
$\textrm{log}_{10} \left( m_{\chi}/\textrm{GeV} \right)$ 		& $(0.1,3.0)$ 	& Uniform prior \\
$\textrm{log}_{10} \left( \sigma_{SI}^p/\textrm{pb}\right)$ 	& $(-10,-6)$ 	& Uniform prior \\
\hline
$\rho_0/(\textrm{GeV/cm}^3)$		& $(0.001,0.9)$	& Gaussian: $0.4 \pm 0.1$\\
$v_0/(\textrm{km/s})$			& $(80,380)$    & Gaussian: $230 \pm 30$  \\
$v_{esc}/(\textrm{km/s})$		& $(379,709)$   & Gaussian: $544 \pm 33$  \\
$k$		  			&  $(0.5,3.5)$	& Uniform prior\\
\hline
\end{tabular}
\caption{\fontsize{9}{9}\selectfont Parameters used in our analysis, with their prior range (middle column) and the prior constraint adopted (rightmost column). See Section \ref{statsection} and \ref{astrounc} for further details.}\label{tabPars}
\end{table*}

The parameter set $\params$ contains the DM quantities we are interested in (mass and scattering cross-section), and also the Galactic model parameters, which we regard as nuisance parameters, entering the calculation of direct detection signals, namely $\rho_0$, $v_0$, $v_{esc}$, $k$, see Eq.~\eqref{recoil2} and Section \ref{astrounc}. We further need to define priors $p(\params)$ for all of our parameters. For the DM parameters, we adopt flat priors on the log of the mass and cross-section, reflecting ignorance on their scale. For the Galactic model parameters, we choose priors that reflect our state of knowledge about their plausible values, as specified in the next section. Those priors are informed by available observational constraints as well as plausible estimations of underlying systematical errors, for example for $\rho_0$. Finally, the likelihood function for each of the direct detection experiments is given by a product of independent Poisson likelihoods over the energy bins: 
\begin{equation}
\like(\params) = \prod_{b} \frac{N_R^{\hat{N}_b}}{\hat{N}_b!}\exp\left(-N_R\right),
\end{equation}
where $\hat{N}_b$ is the number of counts in each bin (generated from the true model with no shot noise, as explained below) and $N_R = N_R(E_b^\text{min}, E_b^\text{max})$ is the number of counts in the $b$-th bin (in the energy range $E_b^\text{min} \leq E \leq E_b^\text{max}$) when the parameters take on the value $\params$, and it is given by Eq.~\eqref{eq:NR}. We use 10 bins for each experiment, uniformly spaced on a linear scale between the threshold energy and 100 keV. We have checked that our results are robust if we double the number of assumed energy bins. Using the experimental capabilities outlined in Section \ref{exp}, we compute the counts $N_R$ that the benchmark WIMPs would generate, and include no background events since the expected background level in the fiducial mass region is negligible (cf.~Table \ref{tabExp}). The mock counts are generated from the true model, i.e.~without Poisson scatter. This is because we want to test the reconstruction capabilities without having to worry about realization noise (such a data set has been called ``Asimov data'' in the particle physics context~\cite{Cowan:2010js}). 

To sample the posterior distribution we employ the MultiNest code~\cite{Feroz:2007kg, Feroz:2008xx,Trotta:2008bp}, an extremely efficient sampler of the posterior
distribution even for likelihood functions defined over a parameter space of large dimensionality with a very complex structure. In our case, the likelihood function is unimodal and well-behaved, so Monte Carlo Markov Chain (MCMC) techniques would be sufficient to explore it. However, MultiNest also computes the Bayesian evidence (which MCMC methods do not return), as it is an implementation of the nested sampling algorithm~\cite{Skilling:2006}. In this work, we run MultiNest with 2000 live points, an efficiency parameter of 1.0 and a tolerance of 0.8 (see \cite{Feroz:2007kg, Feroz:2008xx} for details).

  \begin{figure*}
 \centering
 \includegraphics[width=8.9cm,height=7.6cm]{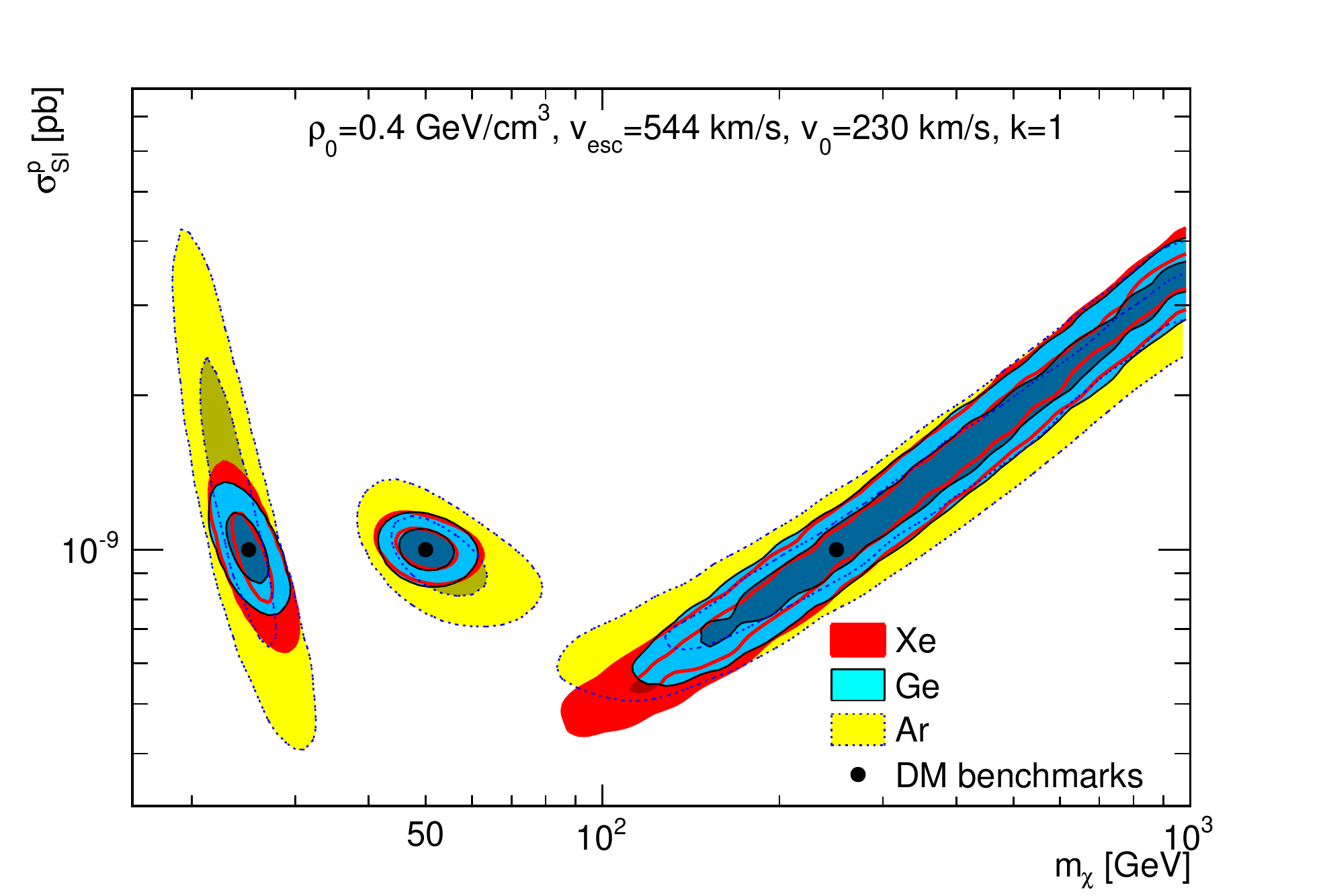}
 \includegraphics[width=8.9cm,height=7.6cm]{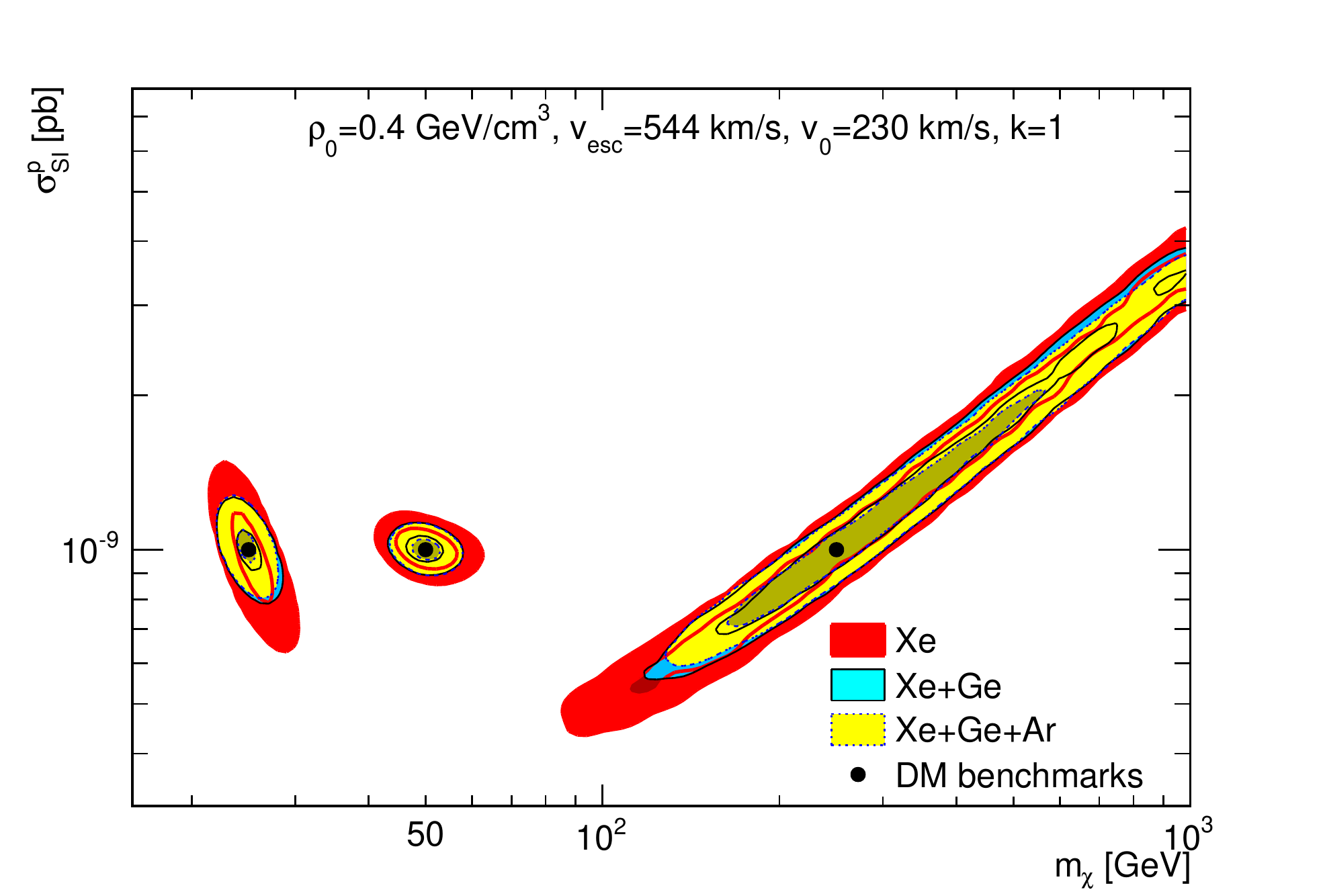}
 \caption{\fontsize{9}{9}\selectfont The joint 68\% and 95\% posterior probability contours in the $m_{\chi}-\sigma_{SI}^p$ plane for the three DM benchmarks ($m_{\chi}=25,50,250$ GeV) with fixed Galactic model, i.e.~fixed astrophysical parameters. In the left frame we show the reconstruction capabilities of Xe, Ge and Ar configurations separately, whereas in the right frame the combined data sets Xe+Ge and Xe+Ge+Ar are shown.}\label{figfixed}
\end{figure*}

\section{Velocity distribution and Galactic model parameters}\label{astrounc}

\par We now move onto discussing our modeling of the velocity distribution function and the Galactic model parameters that are input for Eq.~\eqref{recoil2}. We model only the smooth component of the velocity distribution -- recent results from numerical simulations indicate that the velocity distribution component arising from localised streams and substructures is likely sub-dominant in the calculation of direct dark matter detection signals~\cite{Vogelsberger:2008qb,Kuhlen:2009vh}. 

\par We model the velocity distribution function as spherical and isotropic, and parameterise it as~\cite{Lisanti:2010qx},
\begin{equation}\label{fv}
f(w)=\left\{\begin{array}{ll}
  \frac{1}{N_f}\left[\textrm{exp}\left(\frac{v_{esc}^2-w^2}{k v_0^2}\right)-1\right]^k & \text{if } w \leq v_{esc}\\
  0 & \text{if } w > v_{esc}
\end{array} \right. \, .
\end{equation} 
This velocity distribution function was found to be flexible enough to describe the range of 
dark matter halo profiles found in cosmological simulations~\cite{Lisanti:2010qx}. Boosting into the rest frame of the Earth implies the transformation
$w^2=v^2+v_e^2+2v v_e \textrm{cos}\theta$, where  
$\theta$ is the angle between $\vec{v}$ and $\vec{v}_e\sim \vec{v}_{lsr}$. The shape parameter that determines
the power law tail of the velocity distribution is $k$, the escape velocity is $v_{esc}$, while $v_0$ is a fit parameter 
that we discuss in detail below, and $N_f$ is the appropriate normalisation constant. The special case $k=1$ represents the standard halo model with a truncated Maxwellian distribution, and the corresponding expressions for $N_f$ and $\mathcal{F}$ have been derived analytically in the literature -- see for instance \cite{McCabe}. Note as well that, for any value of $k$, this distribution matches a Maxwellian distribution for sufficiently small velocities $w$ and if $v_{esc}>v_0$.

The high-velocity tail of the distributions found in numerical simulations of pure dark matter galactic halos
are well modelled by $1.5 < k < 3.5$~\cite{Lisanti:2010qx}. In our analysis we will expand this range to also include models that behave similar to pure Maxwellian distributions near the tail of the distribution, so that in our analysis we vary $k$ in the range
\begin{equation}
k = 0.5-3.5 \quad (\textrm{flat}) \quad. 
\end{equation}
We adopt an uniform (i.e., flat) prior within the above range for $k$.

\par The range we take for the $v_{esc}$ is motivated by the results of Ref.~\cite{vescref}, where a sample of high-velocity stars is used to derive a median likelihood local escape velocity of $\bar{v}_{esc}=544$ km/s and a 90\% confidence level interval $498 \textrm{ km/s} < v_{esc} < 608$ km/s.  Assuming Gaussian errors this translates into a 1$\sigma$ uncertainty of 33 km/s.   It is important to note that this constraint on the escape velocity is derived assuming a range in the power law tail for the distribution of stars in the local neighbourhood, which is then related to the power law tail in the dark matter distribution~\cite{vescref}. Motivated by obtaining conservative limits on the reconstructed mass and cross-section of the dark matter, in our modelling we will not include such correlations between the escape velocity and the power law index $k$,
so that in the end we take a Gaussian prior on $v_{esc}$ with mean and standard deviation given by
\begin{equation}
v_{esc}= 544 \pm 33 \textrm{ km/s} \quad (1\sigma) \quad .
\end{equation}

Having specified ranges for $v_{esc}$ and $k$, it remains to consider a range for $v_0$ in Eq.~\eqref{fv}. As defined in that equation, the quantity $v_0$ does not directly correspond to the local circular velocity, $v_{lsr}$, but rather is primarily set by $v_{lsr}$ and the dark matter profile. Following a procedure similar to that discussed in Ref.~\cite{Lisanti:2010qx}, we find the range of values $v_0$ compatible with a given a dark matter halo profile, $\rho_0$ and a range for $v_{lsr}$. For the above range in $v_{lsr}$ and the values $\rho_{0}$ in Eq.~\eqref{eq:rho0} below, we find that the parameter $v_0$ can take values in the range $200-300$ km/s for pure Navarro-Frenk-White (NFW) dark matter halos with outer density slopes $\rho \propto r^{-3}$. Larger values of $v_0$ are allowed for steeper outer density slopes, though the range is found to not expand significantly if we restrict ourselves to models with outer slopes similar to the NFW case. With these caveats in mind regarding the mapping between $v_0$ and $v_{lsr}$ for steeper outer slopes, for simplicity and transparency in our analysis, we will consider a similar range for $v_0$ as for the local circular velocity, so we take $v_0 = v_{lsr}$ (that holds in the case of the standard halo model). 

For the local circular velocity and its uncertainty, a variety of measurements presents a broad range of central values and uncertainties~\cite{v0refs}. To again remain conservative we use an interval bracketing recent determinations:
\begin{equation}
v_0 = v_{lsr} = 230 \pm 30 \textrm{ km/s} \quad (1\sigma) \quad ,
\end{equation}
where we take a Gaussian prior with the above mean and standard deviation. To account for the variation of the local density of dark matter in our modeling, we will take  a mean value and error given by~\cite{CatenaUllio,localDM}
\begin{equation}
\rho_0 = 0.4 \pm 0.1 \textrm{ GeV/cm}^3 \quad (1\sigma) \quad ,
\label{eq:rho0}
\end{equation}
There are several other recent results that determine $\rho_0$, both consistent \cite{Salucci} and somewhat discrepant \cite{Weber:2009pt} 
with our adopted value. Even in light of these uncertainties, we take Eq.~\eqref{eq:rho0} to represent a conservative range for the purposes of our study.

\par For completeness Table \ref{tabPars} summarises the information on the parameters used in our analysis.

\section{Results}\label{results}

\subsection{Complementarity of targets}

\begin{figure*}[t]
 \centering
 \includegraphics[width=8.9cm,height=7.6cm]{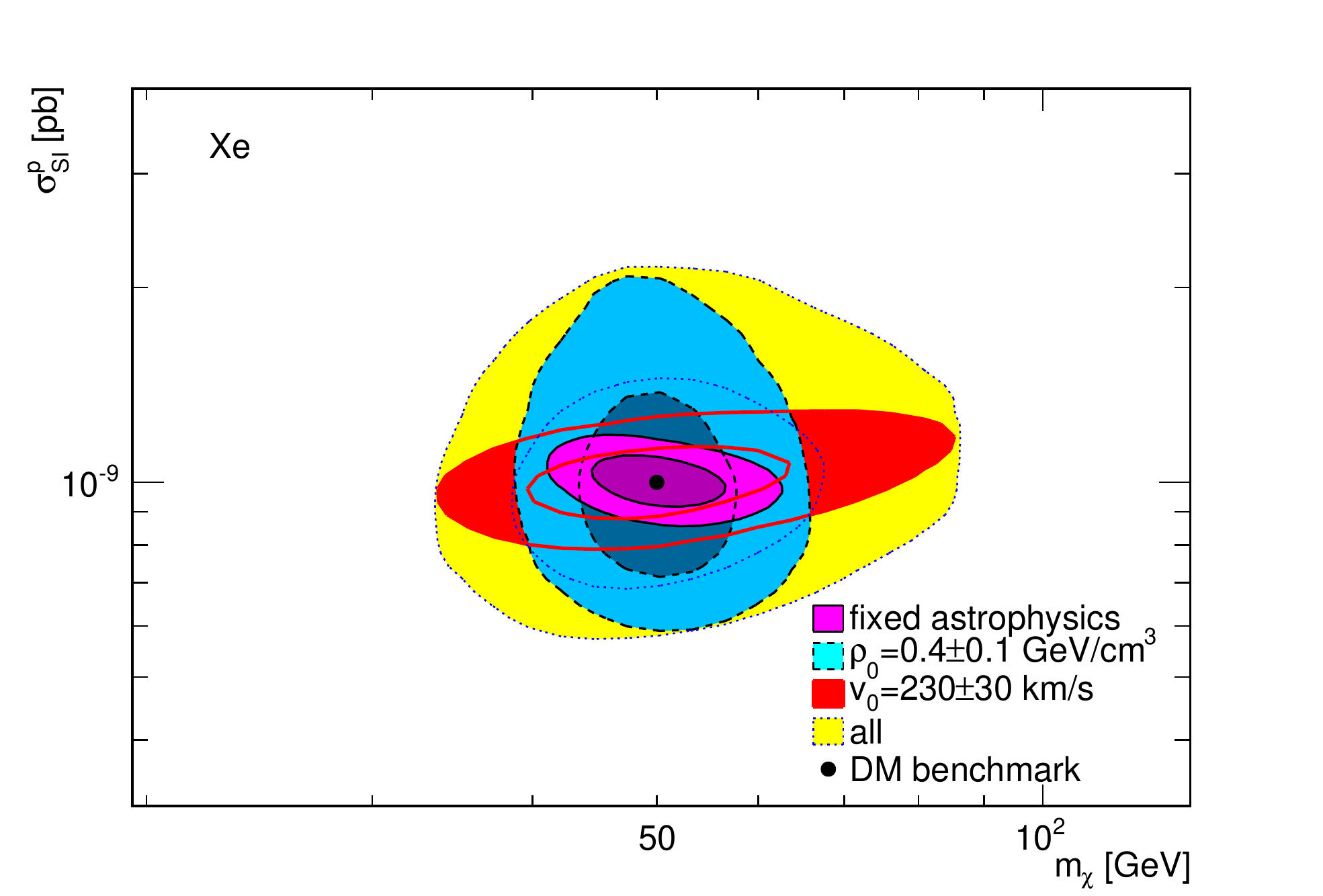}
 \includegraphics[width=8.9cm,height=7.6cm]{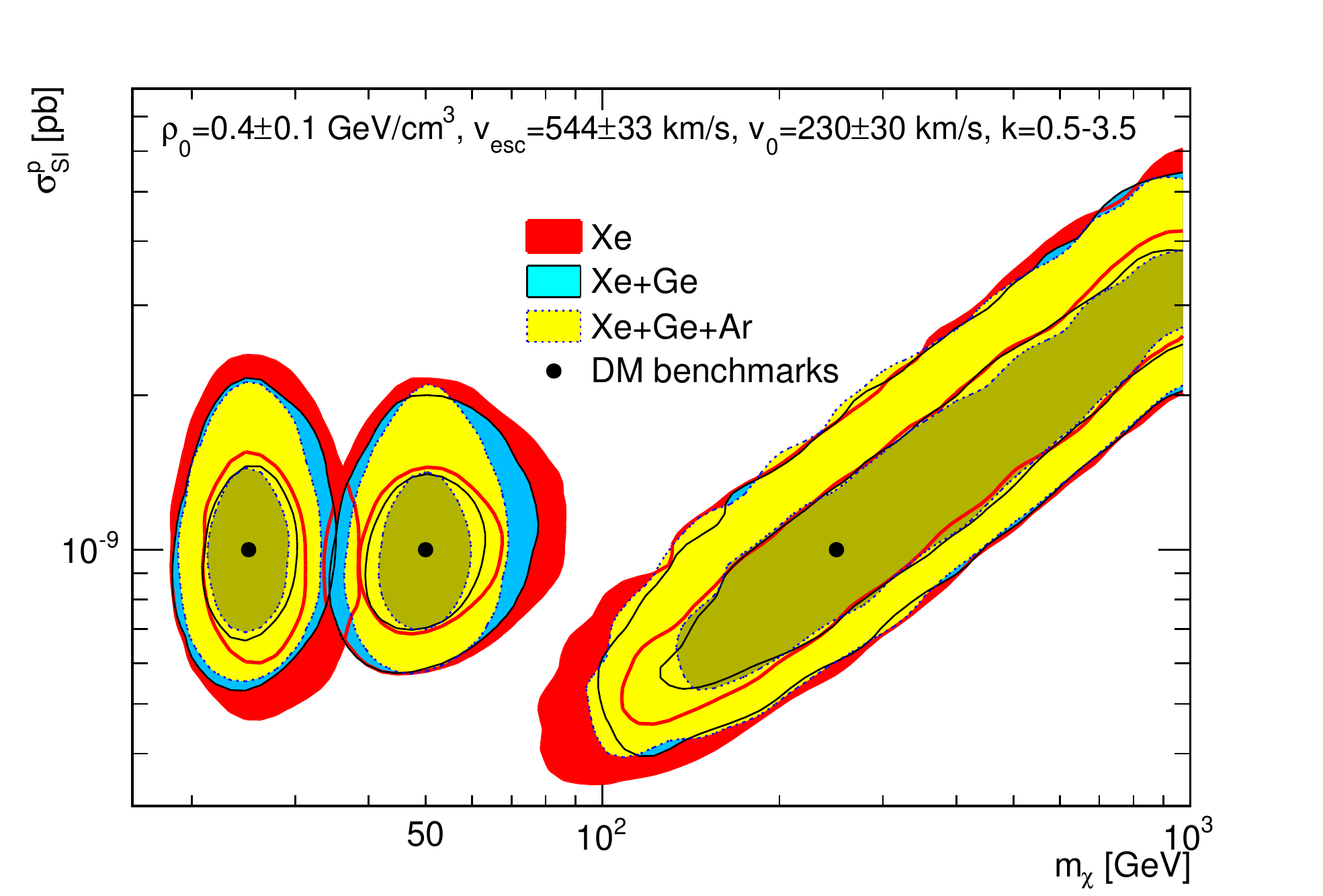}
 \caption{\fontsize{9}{9}\selectfont The joint 68\% and 95\% posterior probability contours in the $m_{\chi}-\sigma_{SI}^p$ plane for the case in which astrophysical uncertainties are taken into account. In the left frame, the effect of marginalising over $\rho_0$, $v_0$ and all four ($\rho_0$, $v_0$, $v_{esc}$, $k$) astrophysical parameters is displayed for a Xe detector and the 50 GeV benchmark WIMP. In the right frame, the combined data sets Xe+Ge and Xe+Ge+Ar are used for the three DM benchmarks ($m_{\chi}=25,50,250$ GeV).}\label{figastro}
\end{figure*}

\par We start by assuming the three dark matter benchmark models described in Section \ref{DD} ($m_{\chi}=25,50,250$ GeV with $\sigma_{SI}^p=10^{-9}$ pb) and fix the Galactic model parameters to their fiducial values, $\rho_0=0.4 \textrm{ GeV}/\textrm{cm}^3$, $v_0=230$ km/s, $v_{esc}=544$ km/s, $k=1$. With the experimental capabilities outlined in Section \ref{exp}, we generate mock data that in turn are used to reconstruct the posterior for the DM parameters $m_{\chi}$ and $\sigma_{SI}^p$. The left frame of Fig.~\ref{figfixed} presents the results for the three benchmarks and for Xe, Ge and Ar separately. Contours in the figure delimit regions of joint 68\% and 95\% posterior probability. Several comments are in order here. First, it is evident that the Ar configuration is less constraining than Xe or Ge ones, which can be traced back to its smaller $A$ and larger $E_{thr}$. Moreover, it is also apparent that, while Ge is the most effective target for the benchmarks with $m_{\chi}=25,250$ GeV, Xe appears the best for a WIMP with $m_{\chi}=50$ GeV (see below for a detailed discussion). Let us stress as well that the 250 GeV WIMP proves very difficult to constrain in terms of mass and cross-section due to the high-mass degeneracy explained in Section \ref{DD}. Taking into account the differences in adopted values and procedures, our results are in qualitative agreement with Ref.~\cite{Akrami1}, where a study on the supersymmetrical framework was performed. However, it is worth noticing that the contours in Ref.~\cite{Akrami1} do not extend to high masses as ours for the 250 GeV benchmark -- this is likely because the volume at high masses in a supersymmetrical parameter space is small.

\begin{table*}
\centering
\fontsize{9}{9}\selectfont
\begin{tabular}{c|c|c}
\hline
& \multicolumn{2}{c}{Percent $1\sigma$ accuracy}\\
  & $m_{\chi}=25$ GeV & $m_{\chi}=50$ GeV  \\
\hline
Xe 	& 6.5\% (14.3\%) 	& 8.1\% (20.4\%) \\
Ge	& 5.5\% (16.0\%)	& 7.0\% (29.6\%)	\\
Ar	& 12.3\% (23.4\%)	& 14.7\% (86.5\%)\\
Xe+Ge	& 3.9\% (10.9\%)	& 5.2\% (15.2\%) \\
Xe+Ge+Ar& 3.6\% (9.0\%)		& 4.5\% (10.7\%)  \\
\hline
\end{tabular}
\caption{\fontsize{9}{9}\selectfont Marginalised percent $1\sigma$ accuracy of the DM mass reconstruction for the benchmarks $m_{\chi}=25,50$ GeV. Figures between brackets refer to scans where the astrophysical parameters were marginalised over (with priors as in Table \ref{tabPars}), while the other figures refer to scans with the fiducial astrophysical setup.}\label{tabAccuracy}
\end{table*}

\par In the right frame of Fig.~\ref{figfixed} we show the reconstruction capabilities attained if one combines Xe and Ge data, or Xe, Ge and Ar together, {\it again for when the Galactic model parameters are kept fixed}. In this case, for $m_{\chi}=25,50$ GeV, the configuration Xe+Ar+Ge allows the extraction of the correct mass to better than $\mathcal{O}(10)$ GeV accuracy. For reference, the (marginalised) mass accuracy for different mock data sets is listed in Table \ref{tabAccuracy}. For $m_{\chi}=250$ GeV, it is only possible to obtain a lower limit on $m_{\chi}$.

Fig.~\ref{figastro} shows the results of a more realistic analysis, that keeps into account the large uncertainties associated with Galactic model parameters, as discussed in Section \ref{astrounc}. The left frame of Fig.~\ref{figastro} shows the effect of varying only $\rho_0$ (dashed lines, blue surfaces), only $v_0$ (solid lines, red surfaces) and all Galactic model parameters (dotted lines, yellow surfaces) for Xe and $m_{\chi}=50$ GeV. The Galactic model uncertainties are dominated by $\rho_0$ and $v_0$, and, once marginalised over, they blow up the constraints obtained with fixed Galactic model parameters. This amounts to a very significant degradation of mass (cf.~Table \ref{tabAccuracy}) and scattering cross-section reconstruction. Inevitably, the complementarity between different targets is affected -- see the right frame of Fig.~\ref{figastro}. Still, for the 50 GeV benchmark, combining Xe, Ge and Ar data improves the mass reconstruction accuracy with respect to the Xe only case, essentially by constraining the high-mass tail.

\begin{figure}[t]
 \centering
 \includegraphics[width=8.9cm,height=7.6cm]{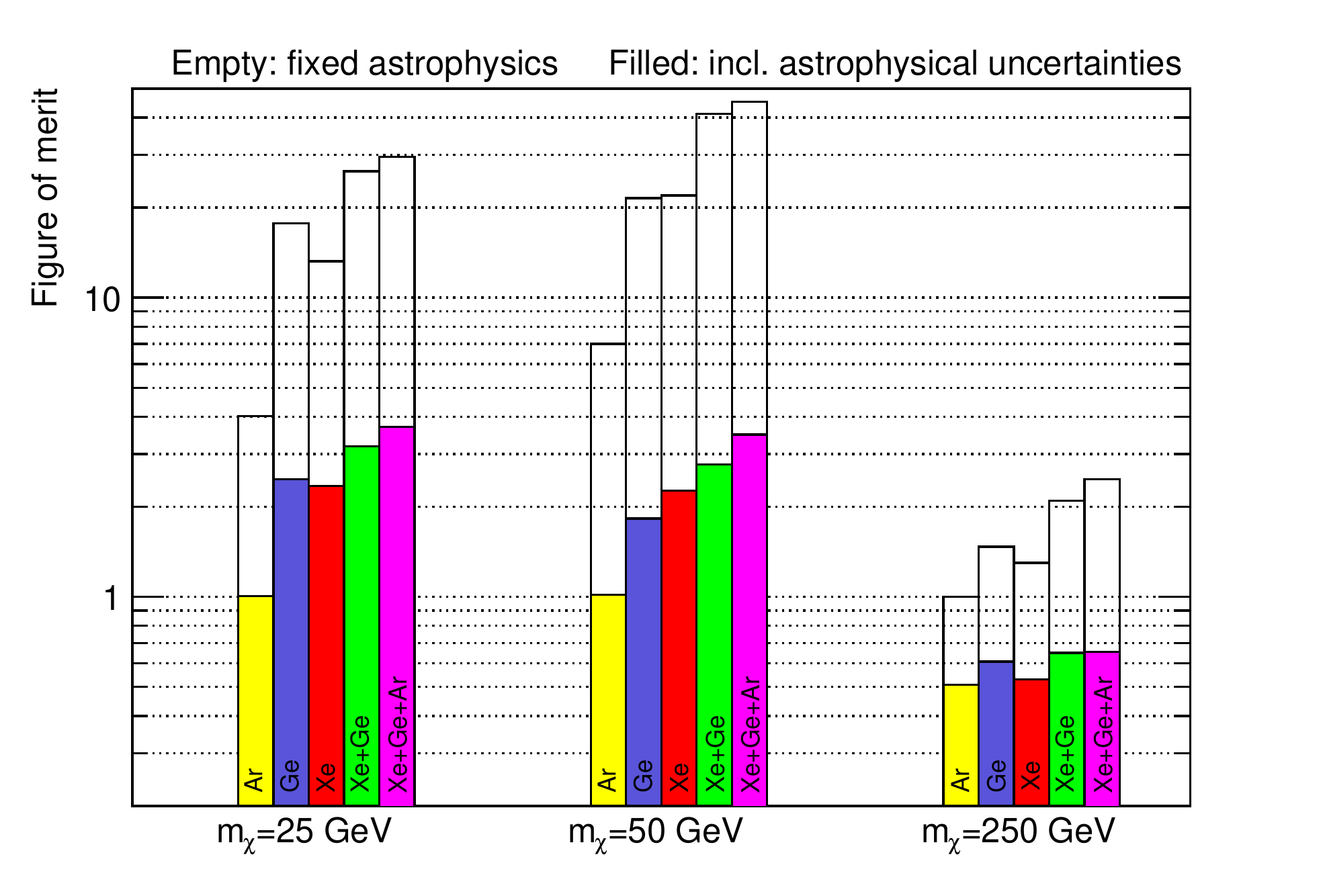}
 \caption{\fontsize{9}{9}\selectfont Figure of merit quantifying the relative information gain on Dark Matter parameters for different targets and combinations thereof. The values of the figure of merit are normalised to the Ar case at $m_{\chi}=250$ GeV with fixed astrophysical parameters. Empty (filled) bars are for fixed astrophysical parameters (including astrophysical uncertainties).}\label{figinfo}
\end{figure}

\par In order to be more quantitative in assessing the usefulness of different targets and their complementarity, we use as {\em figure of merit} the inverse area enclosed by the 95\% marginalised contour in the $\log_{10}(m_{\chi})-\log_{10}(\sigma_{SI}^p)$ plane inside the prior range. Notice that for the 250 GeV benchmark the degeneracy between mass and cross-section is not broken -- this does not lead to a vanishing figure of merit (i.e.~infinite area under the contour) because we are restricting ourselves to the prior range. Fig.~\ref{figinfo} displays this figure of merit for several cases, where we have normalised to the Ar target at $m_{\chi}=250$ GeV with fixed Galactic model. Analyses with fixed Galactic model parameters are represented by empty bars, while the cases where all Galactic model parameters are marginalised over with priors as in Table \ref{tabPars} are represented by filled bars. Firstly, one can see that all three targets perform better for WIMP masses around 50 GeV than 25 or 250 GeV if the Galactic model is fixed. When astrophysical uncertainties are marginalised over, the constraining power of the experiments becomes very similar for benchmark WIMP masses of 25 and 50 GeV. Secondly, Fig.~\ref{figinfo} also confirms what was already apparent from Fig.~\ref{figfixed}: Ge is the best target for $m_{\chi}=25,250$ GeV (although by a narrow margin), whereas Xe appears the most effective for a 50 GeV WIMP (again, by a narrow margin). Furthermore, the inclusion of uncertainties drastically reduces the amount of information one can extract from the data: the filled bars are systematically below the empty ones. Now, astrophysical uncertainties affect the complementarity between different targets in a non-trivial way. To understand this point, let us focus on the two rightmost bars for each benchmark in Fig.~\ref{figinfo}, corresponding to the data sets Xe+Ge and Xe+Ge+Ar. For instance, in the case of a 250 GeV WIMP, astrophysical uncertainties seem to reduce target complementarity: adding Ar to Xe+Ge leads to a significant increase in the figure of merit for analyses with fixed astrophysics (empty bars) but has a negligible effect for analyses with varying astrophysical parameters (filled bars). For low mass benchmarks, the effect of combining two (Xe+Ge) or three targets (Xe+Ge+Ar) is to increase the figure of merit by about a factor of 2 compared to Xe alone or Ge alone, almost independently of whether the astrophysical parameters are fixed or marginalised over. However, the overall information gain on the Dark Matter parameters (for light WIMPs) is reduced by a factor $\sim 10$ if astrophysical uncertainties are taken into account, compared to the case where the Galactic model is fixed.

\begin{table*}[htp]
\centering
\fontsize{8.5}{8.5}\selectfont
\begin{tabular}{c|cccccc|cccccc|cccccc}
\hline
\hline
  & \multicolumn{6}{c|}{$m_{\chi}=25$ GeV} & \multicolumn{6}{c|}{$m_{\chi}=50$ GeV} & \multicolumn{6}{c}{$m_{\chi}=250$ GeV} \\
  & $m_{\chi}$ & $\sigma_{SI}^p$ & $\rho_0$ & $v_0$ & $v_{esc}$ & $k$ & $m_{\chi}$ & $\sigma_{SI}^p$ & $\rho_0$ & $v_0$ & $v_{esc}$ & $k$ & $m_{\chi}$ & $\sigma_{SI}^p$ & $\rho_0$ & $v_0$ & $v_{esc}$ & $k$ \\
\hline
$m_{\chi}$ & $-$ &0.039 & -0.006 & -0.850 & -0.238 & -0.002 & $-$ & 0.098 & -0.006 & -0.870 & -0.079 & -0.004 & $-$ & 0.874 & -0.011 & -0.615 & -0.027 & 0.022 \\

$\sigma_{SI}^p$ & $-$ &	$-$ &	-0.887 & -0.237 & 0.116 & 0.010 & $-$ &	$-$ & -0.957 & -0.175 & 0.026 & -0.031 & $-$ & $-$ &-0.452 & -0.525 & -0.024 & 0.015\\

$\rho_0$ & $-$ & $-$ &	$-$ &	0.013 & -0.005 & 0.005 & $-$ & $-$ & $-$ &	0.014 & -0.010 & 0.030 & $-$ & $-$ & $-$ & 0.002 & 0.015 & 0.010\\

$v_0$ & $-$ &	$-$ &	$-$ &	$-$ &	-0.087 & -0.004 & $-$ & $-$ &$-$ & $-$ &	-0.151 & 0.011 & $-$ & $-$ & $-$ &	$-$ & -0.049 & -0.008 \\

$v_{esc}$ & $-$ & $-$ &	$-$ &	$-$ &	$-$ &	0.000 & $-$ & $-$ &$-$ & $-$ & $-$ & -0.009 & $-$ & $-$ & $-$ &$-$ & $-$ & 0.001\\
\hline
\end{tabular}
\caption{\fontsize{9}{9}\selectfont The correlation factors $r(X,Y)=\text{cov}(X,Y)/(\sigma(X)\sigma(Y))$ for the posteriors obtained from the combined data set Xe+Ge+Ar and including the astrophysical uncertainties with priors as in Table \ref{tabPars}.}\label{tabCorr}
\end{table*}

\subsection{Reduction in uncertainties and self-calibration}

\par
\begin{figure}[t]
 \centering
 \includegraphics[width=8.9cm,height=7.6cm]{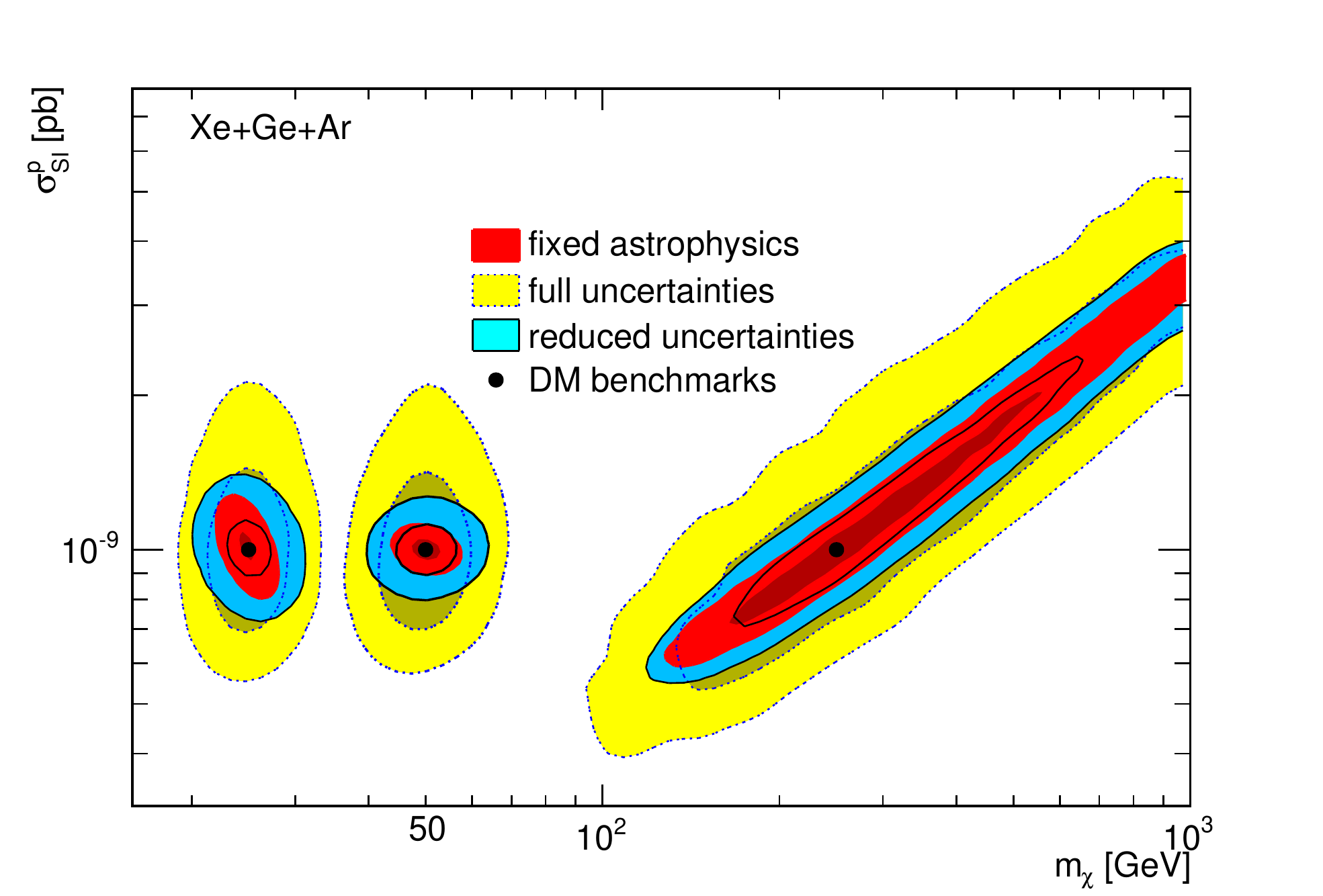}
 \caption{\fontsize{9}{9}\selectfont The effect of reducing the uncertainty on the astrophysical parameters $\rho_0$ and $v_0$. The red surfaces refer to the scan using the fiducial astrophysical setup; the yellow surfaces (and dotted lines) indicate the effect of marginalising over the uncertainties in Table \ref{tabPars}; the blue surfaces (and solid lines) correspond to the reduced uncertainties $\rho_0 = 0.4 \pm 0.028 \textrm{ GeV/cm}^3$, $v_0 = 230 \pm 9.76 \textrm{ km/s}$, $v_{esc}= 544 \pm 33 \textrm{ km/s}$, $k = 0.5-3.5$.}\label{figredastro}
\end{figure}

 The uncertainties used thus far and outlined in Section \ref{astrounc} are a reasonable representation of the current knowledge. For illustration it is also interesting to consider the effect of tighter constraints on Galactic model parameters in the reconstruction of WIMP properties. 
We start by computing the correlation coefficient between the parameters ($m_{\chi}$, $\sigma_{SI}^p$, $\rho_0$, $v_0$, $v_{esc}$, $k$) when they are constrained by the combined data set Xe+Ge+Ar -- see Table \ref{tabCorr}. Clearly, for all benchmark models, $\sigma_{SI}^p$ and $\rho_0$ as well as $m_{\chi}$ and $v_0$ are strongly anti-correlated. The anti-correlation between $\sigma_{SI}^p$ and $\rho_0$ is obvious since $dR/dE_R\propto \sigma_{SI}^p\rho_0$. As for the degeneracy between $m_{\chi}$ and $v_0$, it is easy to verify that, for $v_{min}\ll v_e \sim v_0 \ll v_{esc}$, $\mathcal{F}$ defined in Eq.~\eqref{eq:F} goes approximately as $1/v_0$ and thus $dR/dE_R\propto 1/(m_{\chi}v_0)$. Table \ref{tabCorr} also shows a small (anti-)correlation between $\sigma_{SI}^p$ and $v_0$; all other correlations are negligible. Therefore, $\rho_0$ and $v_0$ are the dominant sources of uncertainty and their more accurate determination will lead to a significant improvement on the reconstruction of $m_{\chi}$ and $\sigma_{SI}^p$. To illustrate this point we follow \cite{CatenaUllio} and apply a 7\% (4.2\%) uncertainty on $\rho_0$ ($v_0$), while maintaining the same central values as before, thus reducing the realistic error bars used above by a factor $\sim 3.0-3.5$ for both parameters. The results are shown in Fig.~\ref{figredastro} where we consider the combination Xe+Ge+Ar. A future more constrained astrophysical setup may indeed lead to a better reconstruction of the WIMP mass and scattering cross-section.

\begin{figure*}
 \centering
 \includegraphics[width=8.9cm,height=7.6cm]{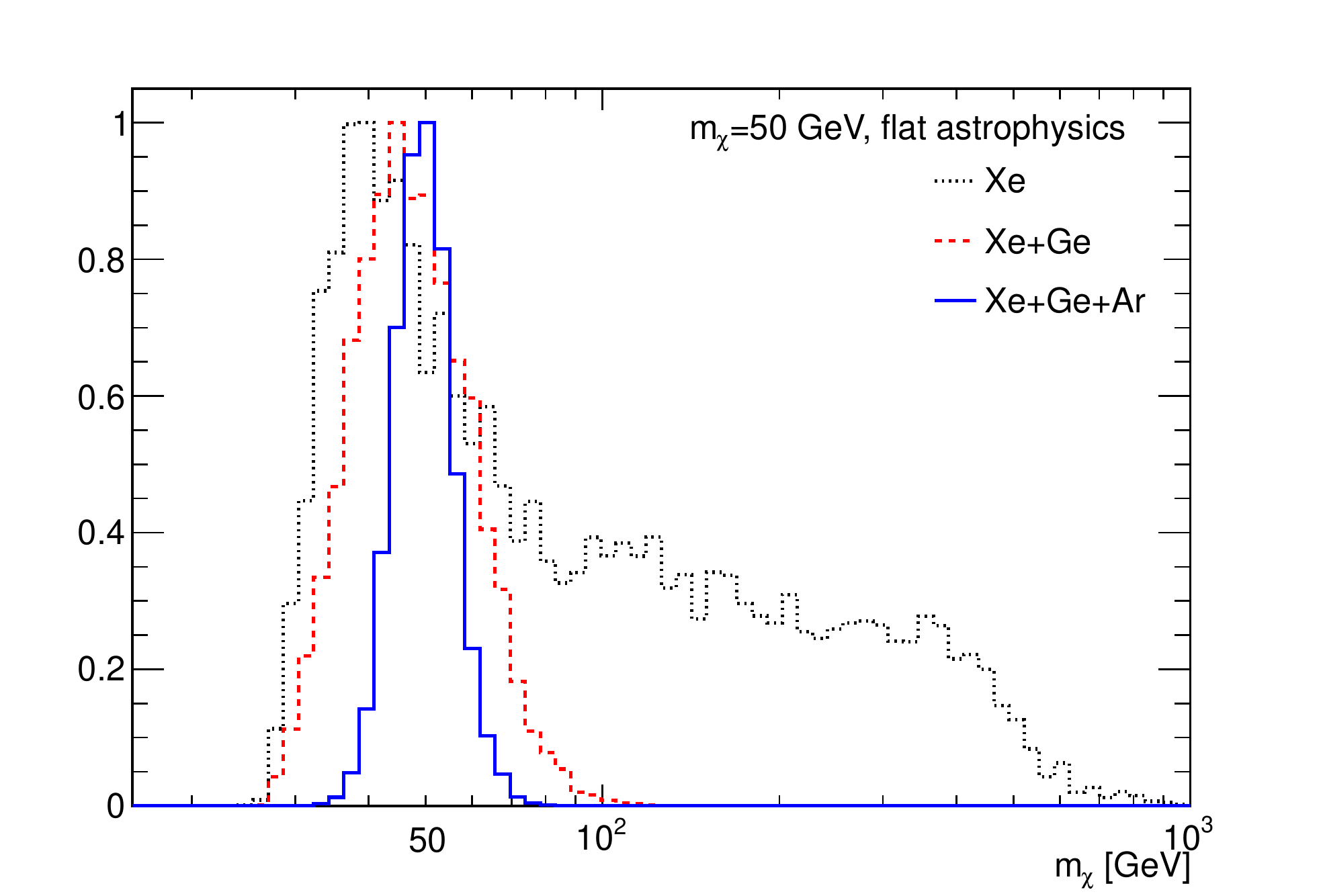}
 \includegraphics[width=8.9cm,height=7.6cm]{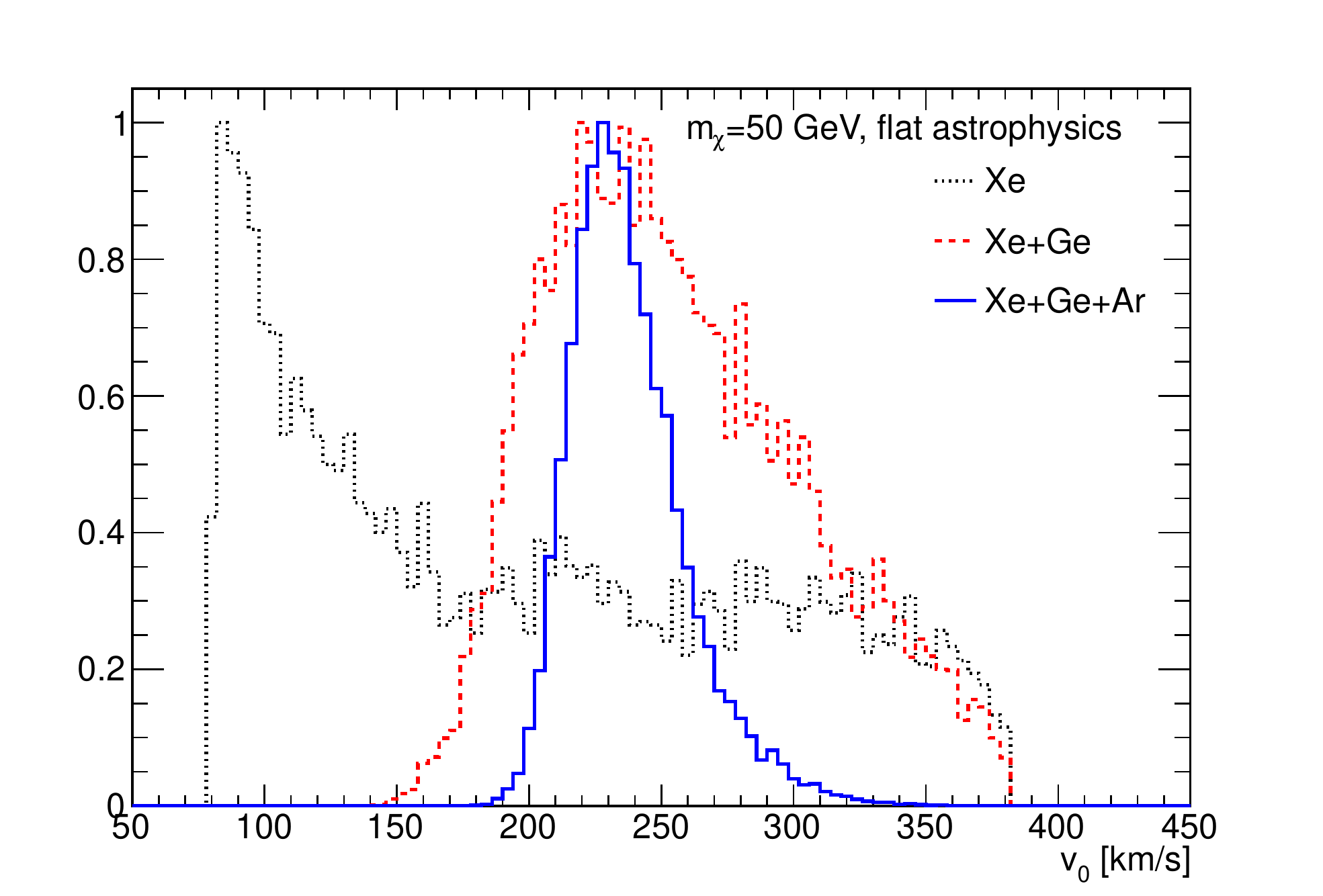}
 \caption{\fontsize{9}{9}\selectfont The marginalised posterior distribution function for $m_{\chi}$ (left frame) and $v_0$ (right frame) with the data sets Xe, Xe+Ge and Xe+Ge+Ar for the 50 GeV benchmark. The parameters $\rho_0$, $v_0$, $v_{esc}$ and $k$ were varied in the ranges indicated in the middle column of Table \ref{tabPars} with a uniform prior and no constraint on astrophysics was applied. The probability distributions are therefore a result of the constraining power of direct detection data only, which have the potential to achieve self-calibration of the circular velocity.}\label{figflatastro}
\end{figure*}

\par To this point we have studied the impact of Galactic model uncertainties on the extraction of DM properties from direct detection data. However, once a positive signal is well-established, it may be used to determine some of the Galactic parameters directly from direct detection data (see e.g.~\cite{APeter}), without relying on external priors. This would amount to achieving a self-calibration of the astrophysical uncertainties affecting direct detection rates. In order to explore such possibility we re-ran our analysis but dropping the Gaussian priors on $\rho_0$, $v_0$ and $v_{esc}$ described in Section \ref{astrounc}. Instead, we used uniform, non-informative priors on $\rho_0$, $v_0$, $v_{esc}$ and $k$ in the ranges indicated in the middle column of Table \ref{tabPars}. We focus on the 50 GeV benchmark and use the data sets Xe, Xe+Ge and Xe+Ge+Ar. With this large freedom on the astrophysical side, it turns out that direct detection data alone leave $\rho_0$, $v_{esc}$ and $k$ unconstrained within their ranges while $\sigma_{SI}^p$ is pinpointed within approximately one order of magnitude. Only the DM mass $m_{\chi}$ and the circular velocity $v_0$ can be constrained by direct detection, as shown in Fig.~\ref{figflatastro}. This figure stresses two interesting results. First, if $m_{\chi}=50$ GeV (and $\sigma_{SI}^p=10^{-9}$ pb), the next generation of experiments will be able to determine the WIMP mass within a few tens of GeV (percent 1$\sigma$ accuracy of 11.8\%) even with very loose assumptions on the local DM distribution. Second, the right frame in Fig.~\ref{figflatastro} shows that the combination of Xe, Ge and Ar targets is very powerful in constraining $v_0$ on its own without external priors. In particular, the data set Xe+Ge+Ar (solid blue line) is sufficient to infer at 1$\sigma$ $v_0=238\pm 22$ km/s (compared to the top-hat prior in the range 80$-$380 km/s). This represents already a smaller uncertainty than the present-day constraint that we have taken, $v_0=230\pm 30$ km/s  -- in case of a positive signal, a combination of direct detection experiments will probe in an effective way the local circular velocity. Repeating the same exercise for the 25 GeV benchmark we find good mass reconstruction but a weaker constraint: $v_0=253\pm 39$ km/s. Again, we stress that the quoted $v_0$ uncertainties in this paragraph do not take into account possible systematic deviations from the parameterisation in Eq. \eqref{fv}.

\section{Conclusions}\label{secconc}

\par We have discussed the reconstruction of the key phenomenological parameters of WIMPs, namely mass and scattering cross-section off nuclei, in case of positive detection with one or more direct DM experiments planned for the next decade. We have in particular studied the complementarity of ton scale experiments with Xe, Ar and Ge targets, adopting experimental configurations that may realistically become available over this time scale. 

To quantify the degree of complementarity of different targets we have introduced a figure of merit measuring the inverse of the area enclosed by the 95\% marginalised contours in the plane $\log_{10}(m_\chi)-\log_{10}(\sigma_{SI}^p)$. There is a high degree of complementarity of different targets: for our benchmark with $m_\chi=50$ GeV and our fiducial set of Galactic model parameters, the relative error on the reconstructed mass goes from 8.1\% for an analysis based on a xenon experiment only, to 5.2\% for a combined analysis with germanium, to 4.5\% adding also argon. Allowing the parameters to vary within the observational uncertainties significantly degrades the reconstruction of the mass, increasing the relative error by up to a factor of $\sim$4 for xenon and germanium, especially due to the uncertainty on $\rho_0$ and $v_0$. However, we found that combining data from Ar, Ge and Xe should allow to reconstruct a 50 GeV WIMP mass to 11.8\% accuracy even under weaker astrophysical constraints than currently available.

Although the mass reconstruction accuracy may appear modest, any improvement of this reconstruction is important, in particular in view of the possible measurement of the same quantity at the Large Hadron Collider at CERN. The existence of a particle with a mass compatible, within the respective uncertainties, with that deduced from direct detection experiments would provide a convincing proof that the particles produced in accelerators are stable over cosmological time scales. Although this is not sufficient to claim discovery of DM \cite{Fornasa}, it would certainly be reassuring.

Despite the strong dependence of direct detection experiments on the Galactic model degrades the reconstruction of DM properties, it does open up the possibility to potentially constrain the local distribution of DM, in case of detection with multiple targets. For example in the case of a low mass 50 GeV WIMP, we have shown that the local circular velocity can be determined from direct detection data alone more accurately than it is presently measured using the local distribution of stars and gas clouds. Additionally, directly detecting DM provides the most realistic way of measuring the local DM velocity distribution. This will in principle provide invaluable information on the structure and formation of the Milky Way halo.

\vspace{0.5cm}
\par {\it Acknowledgements:} G.B., R.T.~and M.P.~would like to thank the organisers of the workshop ``Dark Matter all around'' for a stimulating meeting. We wish to thank the authors of the paper \cite{Akrami1} for providing their preliminary results, as well as Henrique Araujo and Alastair Currie for useful discussions. We also acknowledge support from the SNF grant 20AS21-29329 and the University of Zurich. M.P.~is supported by Funda\c{c}\~{a}o para a Ci\^encia e Tecnologia (Minist\'erio da Ci\^encia, Tecnologia e Ensino Superior).

\end{document}